\documentclass[letter]{aa}  
\usepackage{graphicx}
\usepackage{txfonts}

\def\kms{\ifmmode \mbox{\rm km\ s}^{-1}%   +
        \else {\rm km\ s}$^{-1}$\ignorespaces         % km/s
        \fi}
\def\mum{\ifmmode \mu\mbox{\rm m}  %   +
        \else $\mu${\rm m}         % micro meter
        \fi}
\def\Htwo{\ifmmode \mbox{\rm H}_2   %   +
        \else {\rm H}$_2$         % H2
        \fi}
\def\OHp{\ifmmode \mbox{\rm OH}^+   %   +
        \else {\rm OH$^+$}         % OH+
        \fi}
\def\H2Op{\ifmmode \mbox{\rm H}_2{\rm O}^+   %   +
        \else {\rm H}_2{\rm O}$^+$         % OH+
        \fi}
\def\CHp{\ifmmode \mbox{\rm CH}^+   %   +
        \else {\rm CH}$^+$         % CH+
        \fi}
\def\Hp{\ifmmode \mbox{\rm H}^+   %   +
        \else {\rm H}$^+$         % H+
        \fi}
\def\Hthreep{\ifmmode \mbox{\rm H}^+_3   %   +
        \else {\rm H}$^+_3$         % H3+
        \fi}
\def\mHI{\relax                                      % m
        \ifmmode {\rm m_{\mbox{\scriptsize\rm H\,\sc I}}} %  HI
        \else $m_{\mbox{\scriptsize\rm H\,\sc I}}$
        \fi}
\begin{document}

   \title{Neutral outflows in high-z QSOs}

   \subtitle{}

   \author{Kirsty M. Butler\inst{1,2}
          \and
          Paul P. van der Werf\inst{2}
          \and
          Alain Omont\inst{3}
          \and
          Pierre Cox\inst{3}
          }

   \institute{Institut de Radioastronomie Millimétrique (IRAM), 300 rue de la Piscine, 38400 Saint-Martin-d’Hères, France\\
              \email{butler@iram.fr}
         \and
             Leiden Observatory, Leiden University, PO Box 9513, 2300 RA Leiden, the Netherlands\\
        \and
            Sorbonne Université, UPMC Université Paris 6 and CNRS, UMR 7095, Institut d’Astrophysique de Paris, 98b boulevard Arago, 75014 Paris, France\\
             }

   \date{Received February 28th, 2023; accepted April 20th, 2023 }

% \abstract{}{}{}{}{} 
% 5 {} token are mandatory
 
  \abstract
   {
   \OHp absorption is a powerful tracer of inflowing and outflowing gas in the predominantly atomic diffuse and turbulent halo surrounding galaxies. In this letter, we present observations of $\rm OH^+$($1_1$-$1_0$), CO(9-8) and the underlying dust continuum in five strongly lensed $z\sim2-4$ QSOs, using the Atacama Large Millimeter/submillimeter Array (ALMA) to detect outflowing neutral gas. Blue-shifted $\rm OH^+$($1_1$-$1_0$) absorption is detected in 3/5 QSOs and tentatively detected in a fourth. Absorption at systemic velocities is also detected in one source also displaying blue-shifted absorption. $\rm OH^+$($1_1$-$1_0$) emission is observed in 3/5 QSOs at systemic velocities and the $\rm OH^+$($2_1$-$1_0$) transition is also detected in one source. CO(9-8) is detected in all 5 QSOs at high S/N, providing information on the dense molecular gas within the host galaxy. We compare our sample to high-$z$ far-infrared (FIR) luminous star-forming and active galaxies from the literature. We find no difference in \OHp absorption line properties between active and star-forming galaxies with both samples roughly following the same optical depth-dust temperature relation. This suggests that these observables are driven by the same mechanism in both samples. Similarly, star-forming and active galaxies both follow the same \OHp emission--FIR relation. Obscured QSOs display broader ($>800 \, \kms$) emission than the unobscured QSOs and all but one of the high-z star--forming galaxies (likely caused by the warm molecular gas reservoir obscuring the accreting nucleus). Broader CO(9-8) emission ($>500 \, \kms)$ is found in obscured versus unobscured QSOs, but overall they cover a similar range in line widths as the star-forming galaxies and follow the CO(9-8)--FIR luminosity relation found in low--$z$ galaxies. We find that outflows traced by \OHp are only detected in extreme star-forming galaxies (indicated by broad CO(9-8) emission) and in both types of QSOs, which, in turn, display no red-shifted absorption. This suggests that diffuse neutral outflows in galaxy halos may be associated with the most energetic evolutionary phases leading up to and following the obscured QSO phase.}


   \maketitle
%
%-------------------------------------------------------------------

\section{Introduction}
    Feedback and outflows play a key role in the evolution, regulation, and demise of galaxies throughout cosmic time. Much of the gas accreted onto dark matter halos (and, consequently, their central galaxies, where it condenses to form new stars or feed supermassive black hole growth) is ejected back out of the galaxy via the energetic mechanisms associated with these phenomena. The removal of gas regulates the fuel available for galaxy growth, as well as transporting mass and angular momentum to higher galactic radii \citep{Governato2010}, via fountain flows, or, in more powerful cases, polluting the circumgalactic and intergalactic medium with enriched gas \citep{Travascio2020}. At $z\sim1-3,$ the star formation rate density and black hole accretion peaks in the universe \citep{Madau2014} and, as a result, feedback and outflows must do so as well. 
    
    Outflows are complex multi-phase phenomena, in which the warmer ionised phase is found to dominate the kinetic energy, whilst the cooler neutral and molecular phases dominate the mass and momentum budget of the outflow \citep{Fluetsch2021}. The cooler phases are of particular interest as they remove the direct fuel for star formation, but they have only become available for observation at high--$z$ relatively recently with new and upgraded facilities, such as ALMA and the NOrthern Extended Millimeter Array (NOEMA). 
    
    Low--$z$ studies have commonly made use of high-velocity line wings of bright emission lines to detect outflows \citep{Feruglio2010}, however, at high--$z,$ detecting these weak signals from CO or [C{\small II}] becomes difficult \citep[e.g.][]{Fan2018, Ginolfi2020}. Blue-shifted molecular absorption lines have thus become a popular and reliable way of tracing cool gas outflows both at cosmic noon (e.g. \OHp:\citealt{Butler2021,Riechers2021b,Shao2022}; \CHp: \citealt{Falgarone2017}) and dawn (e.g. OH 119~$\mu m$: \citealt{Spilker2018,Spilker2020b,Butler2023a}; $\rm H_2O$: \citealt{Jones2019}).
    
    One molecule of  note here is $\rm OH^+$, which traces the extended turbulent halo of predominantly atomic and diffuse gas surrounding galaxies \citep{Indriolo2018}. Moreover, the proximity of $\rm OH^+$($1_1$-$1_0$) with the CO(9-8) emission line means that we can simultaneously observe the warm molecular gas, providing additional information on the physical properties within the galaxy \citep{Berta2021,Riechers2021b}. Currently,  observations are limited to star-forming galaxies (\citealt{Butler2021,Riechers2021b,Berta2021,Indriolo2018,Shao2022}, Butler et al. in prep.), with only a few observations achieved in active galaxies \citep{Stanley2021,Shao2019,Shao2022}.
    
    In this letter, we present $\rm OH^+$($1_1$-$1_0$), CO(9-8), and dust continuum observations in five $z{\sim}2{-}4$ far-infrared (FIR) bright QSOs. Throughout this work, we adopt a flat $\Lambda$CDM cosmology with $\Omega_{\rm m}= 0.307$ and ${\rm H_0 = 67.7\ km\ s^{-1}\ Mpc^{-1}}$ \citep{Planck2016}.
%--------------------------------------------------------------------
\section{Sample and observations}
    The data were obtained in the Cycle 7 ALMA project 2019.1.01802.S (P.I.: K.M. Butler), targeting five FIR--bright QSOs at $z\sim2-4$ (Tables \ref{tab:measured} and \ref{tab:derived}). The five quasars were selected based on their 500 \mum continuum flux densities from a sample of 104 gravitationally lensed QSOs presented in \cite{Stacey2018}. The sources are listed in both the CASTLES survey \citep{Kochanek1999} and Sloan Digital Sky Survey Quasar Lens Search catalogue \citep{Inada2012}, which have since been followed up with {\it Herschel}/SPIRE observations \citep{Stacey2018}, providing accurate estimates of their FIR-luminosities and dust temperatures (Table~\ref{tab:derived}). The selected sample covers about two decades in dust temperature, of about order of magnitude in $L_{\rm FIR}$ and including one quasar with a jet-dominated radio emission, MG~J01414+0534 \citep{Stacey2018}. 
    
    All sources were observed with ALMA band 7, except PSS~J2322+1944, which was observed in band 5. The receivers were tuned such that two overlapping spectral windows cover the $\rm OH^+$($1_1$-$1_0$) and CO(9-8) lines in one sideband, with the two remaining spectral windows placed in the second sideband to detect the underlying dust continuum at high S/N (Fig.~\ref{fig:fits}). No calibration issues were found and the observations were all made during good or adequate weather conditions.
    
    The raw data were reduced using CASA \citep{McMullin2007}. The calibrated data were non-interactively imaged using a robust weighting of 0.5 and noise threshold of $1\sigma$ with the \texttt{tclean} routine. We did not subtract the continuum and separated the sidebands into two cubes, leaving the frequency resolution the same as that of the receiver channels (15.624~MHz) (Table \ref{tab:obs}).

%--------------------------------------------------------------------
\section{Spectra and spectral fitting}
    We present both sidebands of the ALMA spectra for each source in Fig.~\ref{fig:fits}. The spectra were created and fitted twice: first by summing over all spaxels with an underlying dust continuum level $\geq 3\sigma$, estimated from a first guess of the line-free channels. We then identified and fit the spectral components and used them to identify any spaxel containing a channel value $\geq 3\sigma$ within the FWHM of one or more of the CO(9-8) components. These spaxels are then included in the spatially integrated spectra and fitted a second time. We fit the spectra with a combination of Gaussian spectral components and a linear continuum slope simultaneously, leaving the central frequencies, line widths, intensities, and continuum gradient as free parameters. We used the same line parameters to describe the $\rm OH^+$($1_1$-$1_0$) and $\rm OH^+$($2_1$-$1_0$) transitions. The final best-fit parameters are presented in Tables~\ref{tab:measured} and \ref{tab:obsWFI}.

\begin{sidewaystable}
    \caption{Observed properties of the dust continuum and the $\rm OH^+(1_1-1_0)$ and CO(9-8) lines.}\label{tab:measured}
    \centering
    \begin{tabular}{lccccccccccc} 
    \cline{1-11}\vspace{-3mm}\\\cline{1-11}\vspace{-3mm}\\    
    Name& \multicolumn{1}{c}{$\rm Cont_{1033.119}$$^*$} & \multicolumn{3}{c}{$\rm OH^+(1_1-1_0)$ Absorption} & \multicolumn{3}{c}{$\rm OH^+(1_1-1_0)$ Emission} & \multicolumn{3}{c}{CO$(9-8)$ Emission}\\
     & S  & $\rm S$&$\rm v$&$\rm \sigma$&$\rm S$&$\rm v$&$\rm \sigma$ & $\rm S$&$\rm v$&$\rm \sigma$\\ %& Grad.
     & $\rm [mJy]$ & $\rm [Jy\ \kms]$ & $\rm [\kms]$ & [\kms] & $\rm [Jy\ \kms]$ & $\rm [\kms]$& [\kms]&[Jy\ \kms] &[\kms] & [\kms]\\ %& $\rm 10^{-14}[Jy\ Hz^{-1}]$ 
    \cline{1-11}\vspace{-3mm}\\   
         HE~1104-1805  & $12.1\pm1.1$  &               &             &              & $1.20\pm0.15$  & $199\pm15$  & $116\pm16$ & $0.89\pm0.13$ & $264\pm3.0$   & $26.2\pm3.0$ \\ %& $12.1\pm3.5$
                       &               &               &             &              &                &             &            & $3.16\pm0.22$  & $154\pm6.5$   & $106\pm6.5$\\
         MG~J0414+0534 & $10.4\pm2.4$  & $-0.73\pm0.17$  & $-162\pm16$ & $144\pm22$   & $4.33\pm0.38$  & $-207\pm49$ & $552\pm54$ & $4.12\pm0.28$  & $-186\pm15$   & $324\pm12$  \\ %& $9.15\pm0.82$
                       &               & $-0.60\pm0.07$   & $-435\pm4.9$& $67.0\pm5.2$ &                &             &            &                &               &     \\
         PSS~J2322+1944& $5.79\pm0.77$ & $-0.25\pm0.13$  & $-308\pm25$ & $84.8\pm32$  & $0.98\pm0.22$ & $-56.0\pm71$& $284\pm48$ & $3.54\pm0.093$ & $-34.7\pm3.4$ & $122\pm3.5$ \\ %& $4.12\pm1.5$ 
         RX~J0911+0551  & $15.9\pm0.74$ & $-0.28\pm0.06$   & $-253\pm21$ & $84.6\pm21$  &                &             &            & $4.14\pm0.36$  & $15.5\pm1.2$  & $42.6\pm0.37$\\ %& $12.4\pm2.0$
                       &               &               &             &              &                &             &            & $1.32\pm0.45$  & $142\pm42$    & $110\pm25$ \\
         WFI~J2026-4536& $7.78\pm2.6$  & $-1.49\pm0.34$ & $-357\pm41$ & $190\pm34$   &                &             &            & $13.1\pm0.31$  & $60.3\pm2.6$  & $108\pm2.8$ \\ %& $17.2\pm8.1$
    \cline{1-11}
    \\
    \multicolumn{12}{l}{\tablefoot{$^*$: Continuum flux at 1033.119 GHz rest frequency (i.e. the $\rm OH^+(1_1-1_0)$ transition rest frame frequency). Uncertainties are taken from those given by the fitting procedure in \texttt{curve\_fit} and may underestimate the \OHp flux errors in the case of MG~J0414+0534 and PSS~J2322+1944 where the line has been broken up into multiple components. Consequently, the deblending of the broad \OHp and CO(9-8) emission lines in MG~J0414+0534 may not be unique. See Fig. \ref{fig:fits} for observed frequencies.}}
\end{tabular}
\end{sidewaystable}

%--------------------------------------------------------------------
\section{Results}
    Here, we present the best-fit parameters of the \OHp absorption, emission and CO(9-8) emission lines in our sample of strongly lensed high-$z$ obscured (MG~J0414+0534) and unobscured (HE~1104-1805, PSS~J2322+1944, RX~J0911+0551 and WFI~J2026-4536) QSOs. We compared our results with high-z sources from the literature with comparable FIR luminosities, including obscured (W0410$-$0913 $z$=3.592 \citealt{Stanley2021}) and unobscured (SDSS~J231038.88+185519.7 $z$=6.0031, \citealt{Shao2022}) QSOs, and DSFGs (HerBS-89a $z$=2.95 \citealt{Berta2021} and the sample of \cite{Riechers2021b}). The central velocity of the CO(9-8) emission (Table \ref{tab:measured}) was used as the systemic velocity when calculating the Doppler--shifted velocities of the \OHp lines.
    
\subsection{Fitted line properties}
\begin{figure*}
   \centering
   \includegraphics[width=\textwidth]{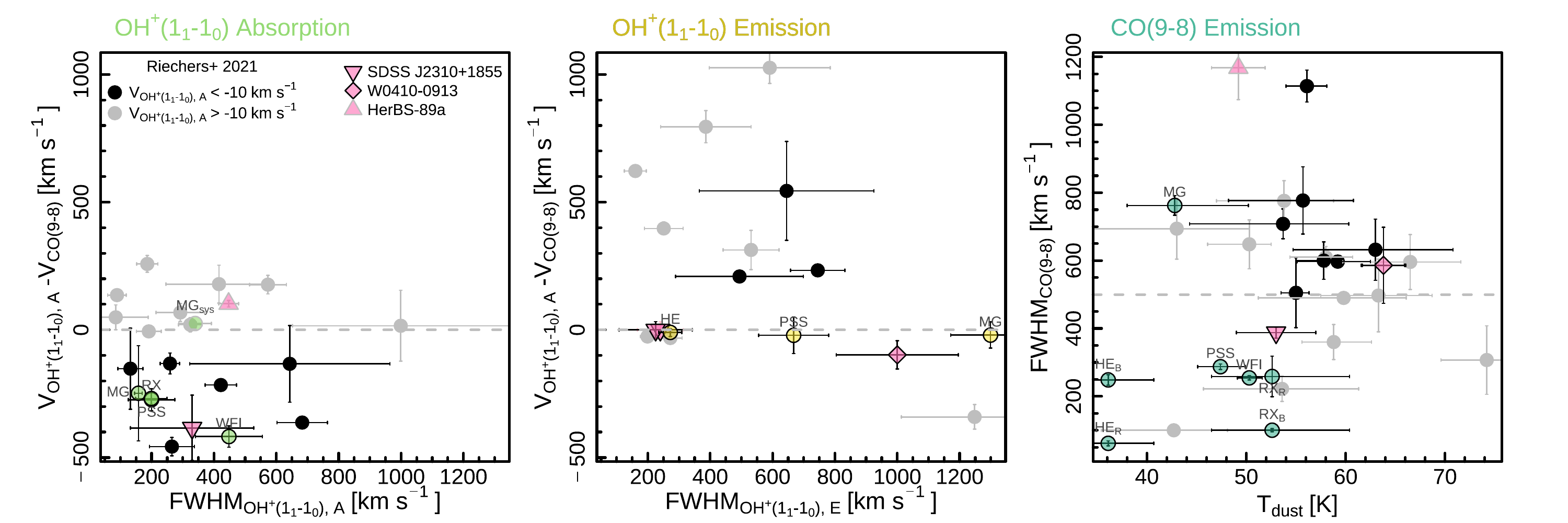}
   \caption{Comparison of the line properties of \OHp and CO(9-8) between the QSO sample of this work with other high--z QSOs \citep{Stanley2021,Shao2022} and DSFGs \citep{Berta2021,Riechers2021b} from the literature. In all panels, we colour the sources from \cite{Riechers2021b} with blue--shifted \OHp absorption in black and colour the rest of the sample in grey. \textbf{Left:} \OHp absorption FWHM vs velocity w.r.t. the CO(9-8) emission. \textbf{Middle:} \OHp emission FWHM vs velocity w.r.t. the CO(9-8) emission. \textbf{Right:} Dust temperature vs CO(9-8) emission FWHM (of only the broadest CO component in each literature source). FWHMs are derived by multiplying the velocity dispersions in Table \ref{tab:measured} by a factor of $2\sqrt{2\log(2)}$. \label{fig:scatter}}
\end{figure*}

    Blue-shifted \OHp absorption is detected in three out of the five QSOs, as well as at systemic velocities in the one obscured and jetted QSO, MG~J0414+0534 (Fig. \ref{fig:fits}, Table \ref{tab:measured}). We include RX~J0911+0551 as a tentative detection as it appears in both the $\rm OH^+$($1_1$-$1_0$) and $\rm OH^+$($2_1$-$1_0$) transitions; however, we stress that these values are uncertain. No red-shifted absorption was found, unlike some of the sources reported in \cite{Riechers2021b} or in the case of HerBS-89a \cite{Berta2021}. Blue-shifted velocities and linewidths are not boosted with respect to the DSFGs. The QSOs show a trend between faster outflow velocity and broader full-width half maximum (FWHM). A larger sample is needed to confirm this finding.
    
    \OHp emission is found in three out of the five QSOs at systemic velocities, unlike in the case of the DSFGs presented in \cite{Riechers2021b}, which display a large spread in velocity offsets between the \OHp and CO(9-8) emission. The obscured QSO MG~J0414+0534 displays the broadest \OHp emission line in the QSO sample.
    
    Strong CO(9-8) emission is observed in all five QSOs. The two obscured QSOs (MG~J0414+0534 and W0410$-$0913) display significantly broader CO(9-8) line widths than the unobscured QSOs. The DSFGs span a wide range of CO(9-8) line widths \citep{Riechers2021b}. However, the eight sources with blue--shifted \OHp absorption all display broad ($\rm FWHM>500\ \kms$) CO(9-8) line widths, similarly to the obscured QSOs and wider than all the unobscured QSOs.
    
\subsection{Derived line properties} 

\begin{table*}
\caption{Derived QSO host galaxy properties and values from the literature.}       
\label{tab:derived}      
\centering          
\begin{tabular}{@{\hskip 3pt}l@{\hskip 1pt}c@{\hskip 1pt}c@{\hskip 1pt}c@{\hskip 2pt}c@{\hskip 2pt}c@{\hskip 3pt}c@{\hskip 2pt}c@{\hskip 2pt}c@{\hskip 2pt}c@{\hskip 3pt}} 
\hline\hline       
Name & $\rm z$ & $\rm T_d$ & $\rm \mu L_{\rm FIR}$$^{a}$ & \multicolumn{2}{c}{ $\rm OH^+(1_1-1_0)$ Absorption} & \multicolumn{2}{c}{$\rm OH^+(1_1-1_0)$ Emission} & \multicolumn{2}{c}{CO(9--8) Emission}\\
 &  & &   & $\rm \int\tau dv$ & $\rm N$ & $\rm \mu L$ & $\rm\mu  L^{\prime}$ & $\rm\mu  L$ & $\rm\mu  L'$\\
 & & [K] & $\rm log_{10}[L_{\odot}]$ &
 [\kms]& $10^{15}[\rm cm^{-2}]$& $\rm 10^{8}[L_{\odot}]$ & $\rm 10^{9}[K\ \kms\ pc^2]$  & $\rm 10^{8}[L_{\odot}]$  & $\rm 10^{9}[K\ \kms\ pc^2]$\\
\hline      
         HE 1104-1805  & $2.3222^{c}$ & 36.1 & $13.22^{+0.04}_{-0.04}$ &                 &        & $1.94\pm0.24$ & $4.03\pm0.51$& $1.44\pm0.22$ & $4.04\pm0.61$ \\
                       &              &      &                         &                 &        &               &              & $3.74\pm0.26$ & $10.5\pm7.4$ \\
         MG J0414+0534 & $2.64^{a}$  & 42.8 & $13.66^{+0.04}_{-0.03}$ & $77.4\pm18$     & $3.77\pm0.87$ & $6.37\pm0.57$ & $18.1\pm1.6$ & $6.09\pm0.41$ & $17.1\pm1.2$ \\
                       &             &      &                         & $70.6\pm8.6$    & $3.44\pm0.42$ &               &              &               &        \\
         PSS J2322+1944& $4.12^{a}$ & 47.4 & $13.58^{+0.01}_{-0.01}$ & $102\pm55$     & $4.99\pm2.7$ & $2.97\pm0.68$ & $8.44\pm1.9$ & $10.8\pm0.28$ & $30.2\pm0.79$ \\
         RX J0911+0551 & $2.79607^{b}$ & 52.6 & $13.58^{+0.04}_{-0.03}$ & $24.3\pm24$     & $1.18\pm 0.27$&               &              & $6.73\pm0.59$ & $18.9\pm1.7$ \\
                       &            &      &                            &                 &        &               &              & $1.32\pm0.45$ & $6.00\pm2.0$ \\
         WFI J2026-4536& $2.24265^{c}$ &  50.3 & $13.81^{+0.02}_{-0.02}$ & $95.6\pm22$    & $4.66\pm1.1$ &               &              & $14.6\pm0.35$ & $41.0\pm0.97$ \\
\hline      
\end{tabular}
\tablefoot{a) FIR luminosities(40–120 \mum) from \cite{Stacey2018}, not corrected for lensing magnification. b) \cite{Stacey2021}, c) This work: we provide updated redshifts for HE~1104$-$1805 of $z=2.3222\pm0.0001$, centered on the $\rm OH^+$($1_1$-$1_0$) emission line, and WFI~J2026$-$4536 $z=2.24265\pm0.00001$, centered on the CO(9-8) emission line.}
\end{table*}

    From the fitted line properties, we derive integrated \OHp absorption optical depths:
        \begin{eqnarray}
        \int\tau\rm dv = -\ln\big(\frac{\rm S_{trans}}{\rm S_{cont}}\big)\rm dv,
        \end{eqnarray}
    where $\rm S_{trans}$ is the transmitted flux and $\rm S_{cont}$ is the unobscured continuum flux level fitted at the central velocity of the line. The \OHp emission and CO(9-8) line luminosities were derived using the expressions given by \cite{Solomon1992} (Table \ref{tab:derived}).
   
\begin{figure*}
   \centering
   \includegraphics[width=\textwidth]{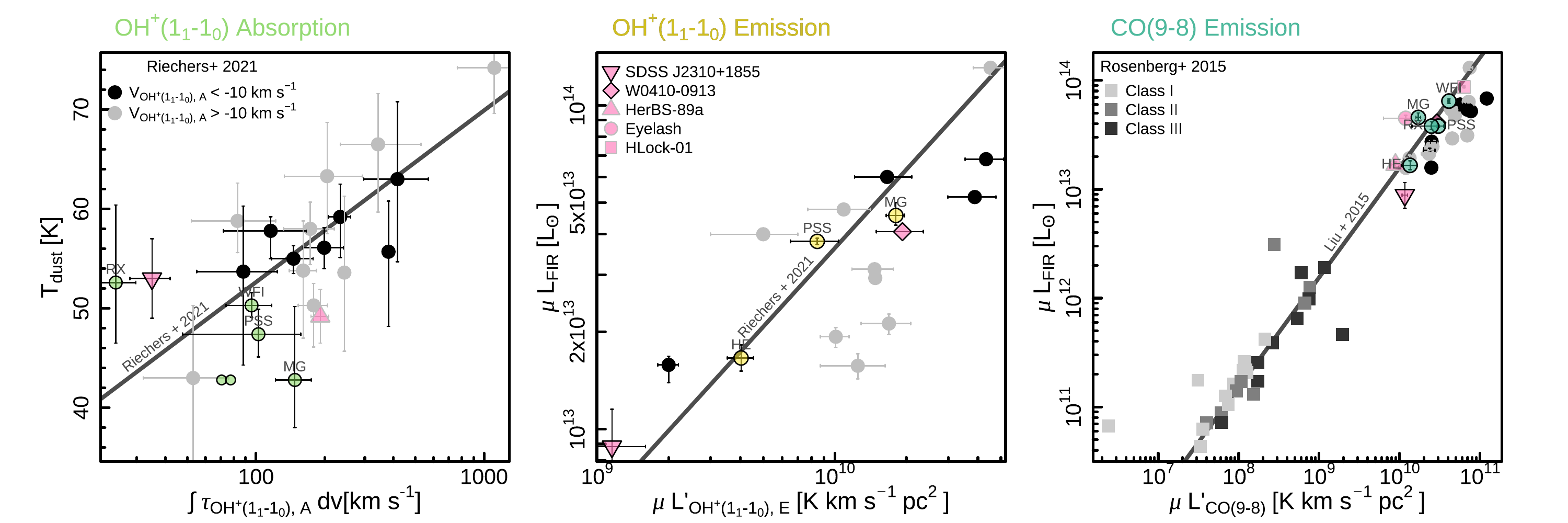}
   \caption{Derived line property trends with host galaxy properties. \textbf{Left:} Integrated \OHp optical depth vs dust temperature. The individual absorption components of MG~J0414+0534 are indicated by smaller green points. \textbf{Middle:} \OHp emission line luminosity vs $\rm L_{\rm FIR}$. \textbf{Right:} CO(9-8) emission line luminosity vs $\rm L_{\rm FIR}$. In this panel, we include the nearby star--forming galaxy sample studied by \cite{Rosenberg2015}, categorised into classes of lowest (class I) to highest (class III) CO excitation. We compare our QSO sample with other high--$z$ QSOs (W0410-0913 \citealt{Stanley2021} and SDSS J2310+1955 \citealt{Shao2022}) and DSFGs (HerBS-89a \citealt{Berta2021}, the Eyelash \citealt{Indriolo2018}, HLock-01 \citealt{Scott2011} and the sample from \citealt{Riechers2021b}) from the literature. In all panels, we colour the sources from \cite{Riechers2021b} with blue--shifted \OHp absorption in black and colour the rest of the sample in grey.\label{fig:derived}}
\end{figure*}

    The integrated \OHp absorption optical depths ($\int\tau_{\rm OH^+, A}$) of the QSO sample lie at the low end of the DSFG sample (Fig.\ref{fig:derived}a), roughly following the $\int\tau_{\rm OH^+, A}$ versus dust temperature relation found by \cite{Riechers2021b}. The QSOs similarly follow the positive\OHp emission line luminosity ($L^\prime_{\rm OH^+, E}$) -- $\rm L_{FIR}$ relation set by the DSFGs. Interestingly, the scatter in this relation is greatly reduced when only considering the DSFGs with detected blue-shifted \OHp absorption.

    Our sample of high-z QSOs follow the $\rm L^\prime_{CO(9-8)}$--$\rm L_{FIR}$ correlation found in low--z galaxies \citep{Liu2015}, with MG~J0414+0534 falling the farthest from the relation towards lower $\rm L^\prime_{\rm CO(9-8)}/L_{\rm FIR}$ ratios (Fig. \ref{fig:derived}c). \cite{Riechers2021b} found that their sample of high-z DSFGs systematically deviates from this trend towards higher $L^\prime_{\rm CO(9-8)}/L_{\rm FIR}$ ratios, a deviation not seen in other high--z star-forming galaxies from the literature.
    
%--------------------------------------------------------------------
\section{Discussion}
\subsection{\OHp Absorption}
$\rm OH^+$($1_1$-$1_0$) absorption has proven to be a reliable tracer of turbulent, diffuse, and predominantly atomic gas surrounding galaxies \citep[e.g.][]{Indriolo2018} revealing both inflowing \citep{Berta2021,Riechers2021b} and outflowing gas moving through the CGM at high--$z$ \citep{Indriolo2018,Butler2021,Riechers2021b,Shao2022}.
\OHp absorption and emission are detected at similar rates ($\sim75\%$ and $\sim65\%$, respectively) in the high--$z$ DSFGs and QSOs samples; however, \cite{Riechers2021b} reported similar numbers of red- and blue-shifted \OHp absorption in their DSFG sample, although current numbers of high-$z$ sources showing clear evidence of infalling gas remain sparse \citep[][and references therein]{Berta2021, Riechers2021b}. We did not find any occurrences of red-shifted absorption (Fig.~\ref{fig:scatter}), however, our selection of mostly Type 1 AGN systems (4/5 sources) may bias our results, as gas is more likely to infall perpendicularly to the opening of the active nucleus, where counter-acting radiation can escape most efficiently from the host galaxy \cite[see, e.g.][]{Shao2022}. Furthermore, we cannot rule out red--shifted absorption in WFI J2026-4536, where this frequency range is not covered. Alternatively, this could be a real difference between DSFGs and QSOs, indicating that feeding from the CGM has been suppressed or halted by feedback in QSO hosts. The \OHp absorption in the QSO sample does not otherwise display faster or broader lines (Fig.~\ref{fig:scatter}) and approximately follows the relation in optical depth with dust temperature set by the DSFGs (Fig.~\ref{fig:derived}). This suggests that the ejection of gas traced by \OHp absorption is not significantly affected by the presence of an active galactic nucleus, nor by whether the nucleus is obscured or not. 
    
\subsection{\OHp Emission}
    In emission, \OHp traces environments with high electron density (e.g. \citealt{Gerin2016}), which can arise in the dense hot gas in both photon- and X-ray-dominated regions (PDRs, XDRs) and has thus been detected in both active \citep{vanderWerf2010,Stanley2021,Shao2022} and star-forming galaxies (\citealt{Stanley2021,Riechers2021b}, Butler et al., in prep.).
    
    The two obscured AGNs, MG~J0414+0534 (see Fig.~\ref{fig:fits}) and W0410$-$0913 \citep{Stanley2021}, display very broad ($>1000$ \kms) \OHp emission, with only one source in the DSFG sample of \cite{Riechers2021b} with a comparable linewidth. The remaining DSFGs and unobscured QSOs lie in similar ranges. A greater contribution from XDRs, a higher prevalence of both XDRs and PDRs or the presence of a wind component in the obscuring molecular gas reservoir directly surrounding the active nucleus in obscured QSOs may be responsible for their broader emission lines. We did not see a boost in the \OHp emission line luminosity with respect to the $L_{\rm FIR}$ of any of the QSOs,  instead finding good agreement with the DSFG relation. Considerable AGN contributions to the $\rm L_{\rm FIR}$ may be expected in the obscured active systems \citep{Schneider2015,Duras2017}, however, MG~J0414+0534 and W0410$-$0913 do not display n offset towards lower $L^\prime_{\rm OH^+}/L_{\rm FIR}$ ratios. This may indicate that the central QSO is also contributing to the $\rm L^\prime_{\rm OH^+}$, such that the $L^\prime_{\rm OH^+}/L_{\rm FIR}$ ratio is maintained.
    
\subsection{CO(9-8) Emission}
    CO(9-8) is predominantly excited by mechanisms associated with warm dense molecular gas in star-forming regions. AGN can contribute to the CO(9-8) emission when present but typically do not dominate until higher J transitions \citep[e.g.][]{Li2020}. 

    Whilst overall the QSO and DSFG samples cover similar ranges in CO(9-8) line width, the obscured systems (MG~J0414+0534 and W0410-0913 \citealt{Stanley2021}) show significantly broader CO(9-8) emission than the unobscured QSOs. This is in agreement with \cite{Stacey2022}, who showed that red, obscured QSOs (including MG~J0414+0534) display broader ($\gtrsim500$ \kms) high-J CO lines than their unobscured counterparts. Comparing high-J line widths with that of bulk gas tracers (i.e. low-J CO transitions or [CI]), they show that the high-J transitions in reddened sources displayed excess flux at high velocities. They attribute this emission to molecular gas winds driven by radiation pressure trapped by the obscuring material around the active nuclei. Narrow CO(9-8) emission observed in unobscured QSOs thus indicates a phase after which the obscuring material has been ejected and radiation from the central AGN can efficiently escape. 

    Blue-shifted \OHp absorption is detected in QSOs displaying both broad and narrow CO(9-8) emission but only in DSFGs displaying broad CO(9-8) emission. This may simply be due to the higher S/N of the QSO spectra, thus requiring larger samples at high S/N to be confirmed. If confirmed, this may indicate that neutral outflows in galaxy halos require extreme galaxies, namely, those displaying broad CO(9-8) emission or harbouring an AGN). Following the evolutionary picture where heavily star--forming galaxies evolve into quiescent galaxies via a short-lived QSO phase \citep{Simpson2012}, blue-shifted \OHp absorption may be indicative of the energetic phases building up to an obscured QSO, and into the unobscured QSO phase \citep{Petter2023}. This is consistent with the absence of red-shifted \OHp absorption in the QSO sample. Larger samples targeting QSOs at all evolutionary stages are needed to test this hypothesis.
    
    \cite{Riechers2021b} suggest a higher prevalence of shock excitation causes the systematic deviation of their high--$z$ DSFG sample from the low--z $L^\prime_{\rm CO(9-8)}-L_{\rm FIR}$ relation. At low--$z$, sources categorised into Class I, II, and III in order of increasing CO excitation, showed trends of falling above, on, and both above and below the relation with greater scatter, respectively (\citealt{Rosenberg2015}, Fig.\ref{fig:derived}). \cite{Riechers2021b} noted that their sample fall into a similar offset range as half the Class III sources located below the relation. Additionally, the highest $\rm L^\prime_{\rm CO(9-8)}/L_{\rm FIR}$ ratio was found in the binary active galactic nucleus NGC~6240, believed to be driven by shock excitation \citep{Meijerink2013}. Also citing \OHp emission as an indicator of shocks, \cite{Riechers2021b} suggested that shock heating drives the shift towards higher $\rm L^\prime_{\rm CO(9-8)}/L_{\rm FIR}$ ratios in their sample. 
    
    With the presence of both intense star formation and an AGN, shocks may be expected to be even more prevalent in the high-$z$ QSO sample, particularly in the obscured sources. MG~J0414+0534, however, displays the lowest $\rm L^\prime_{\rm CO(9-8)}/L_{\rm FIR}$ ratio in our sample with the other high-z sources from the literature, falling tightly on the low-$z$ relation. Furthermore, \OHp emission traces regions of high electron density (PDRs/XDRs) but not necessarily shocked gas and, in any case, the QSOs follow the same $\rm L^\prime_{\rm OH^+}-L_{\rm FIR}$ trend found by \cite{Riechers2021b}, indicating that the deviation between samples in the $\rm L^\prime_{\rm CO(9-8)}-L_{\rm FIR}$ plane is unrelated.
    
    AGN contributions to the $\rm L_{\rm FIR}$ may be expected in the QSO sample, accounting for the upwards vertical offset from the DSFG sample in Fig.~\ref{fig:derived}. Considerable evidence for this is found in obscured QSOs \citep{Duras2017,Schneider2015}, however, studies of unobscured QSOs at high-$z$ do not find evidence for significant AGN contributions \citep{Venemans2018,Venemans2020,Butler2023a}. Furthermore, this deviation is not found between our sample and the high-$z$ DSFGs from the literature that were not studied by \cite{Riechers2021b}.  Therefore, we suggest that the systematic deviation of the DSFGs studied by \cite{Riechers2021b} towards higher $\rm L^\prime_{\rm CO(9-8)}/L_{\rm FIR}$ ratios is a characteristic of that sample and not representative of high-$z$ DSFGs in general.

\subsection{Spatial offsets and differential lensing}
    Differing spatial distributions between separate components (i.e. gas and dust) in a gravitationally lensed source will result in differential lensing. Whilst optical depth, central velocity, and line width are not impacted by gravitational lensing, the relative magnification of line and dust emission is; this leads to the shifting of sources seen in Fig. \ref{fig:derived}c,d. The S/N of our observations are not sufficient to reliably disentangle the blended emission and absorption lines on a spaxel-by-spaxel base, however, the sample's agreement with unlensed low-z FIR scaling relations suggests that differential lensing is not substantial. Furthermore, the majority of the high-z comparison sample is strongly lensed, thus making differential lensing an unlikely cause of the offset found in the sample of \cite{Riechers2021b}.  Therefore, we do not believe differential lensing has significantly impacted the conclusions presented below for this study.

%--------------------------------------------------------------------
\section{Conclusions}
    We present ALMA observations targeting the \OHp($1_1-1_0$) and CO(9-8) transitions in five $z\sim 2-4$ far-infrared luminous QSOs. The \OHp($1_1-1_0$) absorption and emission are seen at similar rates in star-forming and active galaxies at high--$z$. However, \OHp absorption is found only at systemic or blue-shifted velocities in the QSO sample, unlike in DSFGs, where red-shifted absorption has also been detected \citep{Berta2021, Riechers2021b} -- although red--shifted absorption cannot be ruled out in one QSO, WFI J2026-4536. We do not find significant differences in the \OHp absorption line properties between active and star-forming samples, suggesting that the central QSO plays a minor role in the ejection of diffuse gas in the halo but may still stop or suppress inflowing gas from the CGM.

    Obscured QSOs display broader \OHp emission lines compared with unobscured QSOs and DSFGs, likely due to a higher prevalence of XDRs, PDRS, or molecular winds in the obscuring reservoir. Obscured QSOs similarly display broader CO(9-8) emission than unobscured sources, likely due to massive molecular outflows typical of the blow-out phase currently underway in the obscured QSOs \citep{Stacey2022}. Despite differences found in the emission lines of the obscured vs unobscured QSOs, we find no differences in the outflow properties traced by \OHp absorption but we do note that only DSFGs with ($\rm FWHM_{\rm CO(9-8)}>500$ \kms) CO(9-8) emission have blue-shifted \OHp absorption detected. This may indicate that diffuse, neutral outflows in the CGM are driven by the most extreme sources (i.e. AGNs or displaying broad emission lines). We therefore suggest that blue-shifted \OHp absorption may be indicative of the energetic phases leading up to the obscured QSO phase and into the unobscured phase where infalling gas has been halted (red-shifted absorption).
    
\begin{acknowledgements}
      The Authors would like the thank the anonomous referee who helped improve this letter. This paper makes use of the following ALMA data: ADS/JAO.ALMA\#2019.1.01802.S. ALMA is a partnership of ESO (representing its member states), NSF (USA) and NINS (Japan), together with NRC (Canada), MOST and ASIAA (Taiwan), and KASI (Republic of Korea), in cooperation with the Republic of Chile. The Joint ALMA Observatory is operated by ESO, AUI/NRAO and NAOJ. This work benefited from the support of the project Z-GAL ANR-AAPG2019 of the French National Research Agency (ANR).
\end{acknowledgements}
%TC:endignore

\bibliographystyle{aa.bst} % style aa.bst
\bibliography{OHpz2QSOs.bib} % your references Yourfile.bib

\begin{thebibliography}{40}
\expandafter\ifx\csname natexlab\endcsname\relax\def\natexlab#1{#1}\fi

\bibitem[{{Berta} {et~al.}(2021){Berta}, {Young}, {Cox}, {Neri}, {Jones},
  {Baker}, {Omont}, {Dunne}, {Carnero Rosell}, {Marchetti}, {Negrello}, {Yang},
  {Riechers}, {Dannerbauer}, {Perez-Fournon}, {van der Werf}, {Bakx}, {Ivison},
  {Beelen}, {Buat}, {Cooray}, {Cortzen}, {Dye}, {Eales}, {Gavazzi}, {Harris},
  {Herrera}, {Hughes}, {Jin}, {Krips}, {Lagache}, {Lehnert}, {Messias},
  {Serjeant}, {Stanley}, {Urquhart}, {Vlahakis}, \& {Wei{\ss}}}]{Berta2021}
{Berta}, S., {Young}, A.~J., {Cox}, P., {et~al.} 2021, \aap, 646, A122

\bibitem[{{Butler} {et~al.}(2021){Butler}, {van der Werf}, {Rybak}, {Costa},
  {Cox}, {Wei{\ss}}, {Micha{\l}owski}, {Riechers}, {Rigopoulou}, {Marchetti},
  {Eales}, \& {Valtchanov}}]{Butler2021}
{Butler}, K.~M., {van der Werf}, P.~P., {Rybak}, M., {et~al.} 2021, \apj, 919,
  5

\bibitem[{{Butler} {et~al.}(2023){Butler}, {van der Werf}, {Topkaras}, {Rybak},
  {Venemans}, {Walter}, \& {Decarli}}]{Butler2023a}
{Butler}, K.~M., {van der Werf}, P.~P., {Topkaras}, T., {et~al.} 2023, \apj,
  944, 134

\bibitem[{{Duras} {et~al.}(2017){Duras}, {Bongiorno}, {Piconcelli}, {Bianchi},
  {Pappalardo}, {Valiante}, {Bischetti}, {Feruglio}, {Martocchia}, {Schneider},
  {Vietri}, {Vignali}, {Zappacosta}, {La Franca}, \& {Fiore}}]{Duras2017}
{Duras}, F., {Bongiorno}, A., {Piconcelli}, E., {et~al.} 2017, \aap, 604, A67

\bibitem[{{Falgarone} {et~al.}(2017){Falgarone}, {Zwaan}, {Godard}, {Bergin},
  {Ivison}, {Andreani}, {Bournaud}, {Bussmann}, {Elbaz}, {Omont}, {Oteo}, \&
  {Walter}}]{Falgarone2017}
{Falgarone}, E., {Zwaan}, M.~A., {Godard}, B., {et~al.} 2017, \nat, 548, 430

\bibitem[{{Fan} {et~al.}(2018){Fan}, {Knudsen}, {Fogasy}, \&
  {Drouart}}]{Fan2018}
{Fan}, L., {Knudsen}, K.~K., {Fogasy}, J., \& {Drouart}, G. 2018, \apjl, 856,
  L5

\bibitem[{{Feruglio} {et~al.}(2010){Feruglio}, {Maiolino}, {Piconcelli},
  {Menci}, {Aussel}, {Lamastra}, \& {Fiore}}]{Feruglio2010}
{Feruglio}, C., {Maiolino}, R., {Piconcelli}, E., {et~al.} 2010, \aap, 518,
  L155

\bibitem[{{Fluetsch} {et~al.}(2021){Fluetsch}, {Maiolino}, {Carniani},
  {Arribas}, {Belfiore}, {Bellocchi}, {Cazzoli}, {Cicone}, {Cresci}, {Fabian},
  {Gallagher}, {Ishibashi}, {Mannucci}, {Marconi}, {Perna}, {Sturm}, \&
  {Venturi}}]{Fluetsch2021}
{Fluetsch}, A., {Maiolino}, R., {Carniani}, S., {et~al.} 2021, \mnras, 505,
  5753

\bibitem[{{Gerin} {et~al.}(2016){Gerin}, {Neufeld}, \&
  {Goicoechea}}]{Gerin2016}
{Gerin}, M., {Neufeld}, D.~A., \& {Goicoechea}, J.~R. 2016, \araa, 54, 181

\bibitem[{{Ginolfi} {et~al.}(2020){Ginolfi}, {Jones}, {B{\'e}thermin},
  {Fudamoto}, {Loiacono}, {Fujimoto}, {Le F{\'e}vre}, {Faisst}, {Schaerer},
  {Cassata}, {Silverman}, {Yan}, {Capak}, {Bardelli}, {Boquien}, {Carraro},
  {Dessauges-Zavadsky}, {Giavalisco}, {Gruppioni}, {Ibar}, {Khusanova},
  {Lemaux}, {Maiolino}, {Narayanan}, {Oesch}, {Pozzi}, {Rodighiero}, {Talia},
  {Toft}, {Vallini}, {Vergani}, \& {Zamorani}}]{Ginolfi2020}
{Ginolfi}, M., {Jones}, G.~C., {B{\'e}thermin}, M., {et~al.} 2020, \aap, 633,
  A90

\bibitem[{{Governato} {et~al.}(2010){Governato}, {Brook}, {Mayer}, {Brooks},
  {Rhee}, {Wadsley}, {Jonsson}, {Willman}, {Stinson}, {Quinn}, \&
  {Madau}}]{Governato2010}
{Governato}, F., {Brook}, C., {Mayer}, L., {et~al.} 2010, \nat, 463, 203

\bibitem[{{Inada} {et~al.}(2012){Inada}, {Oguri}, {Shin}, {Kayo}, {Strauss},
  {Morokuma}, {Rusu}, {Fukugita}, {Kochanek}, {Richards}, {Schneider}, {York},
  {Bahcall}, {Frieman}, {Hall}, \& {White}}]{Inada2012}
{Inada}, N., {Oguri}, M., {Shin}, M.-S., {et~al.} 2012, \aj, 143, 119

\bibitem[{{Indriolo} {et~al.}(2018){Indriolo}, {Bergin}, {Falgarone}, {Godard},
  {Zwaan}, {Neufeld}, \& {Wolfire}}]{Indriolo2018}
{Indriolo}, N., {Bergin}, E.~A., {Falgarone}, E., {et~al.} 2018, \apj, 865, 127

\bibitem[{{Jones} {et~al.}(2019){Jones}, {Maiolino}, {Caselli}, \&
  {Carniani}}]{Jones2019}
{Jones}, G.~C., {Maiolino}, R., {Caselli}, P., \& {Carniani}, S. 2019, \aap,
  632, L7

\bibitem[{{Kochanek} {et~al.}(1999){Kochanek}, {Falco}, {Impey}, {Leh{\'a}r},
  {McLeod}, \& {Rix}}]{Kochanek1999}
{Kochanek}, C.~S., {Falco}, E.~E., {Impey}, C.~D., {et~al.} 1999, in American
  Institute of Physics Conference Series, Vol. 470, After the Dark Ages: When
  Galaxies were Young (the Universe at  Z 2-5), ed. S.~{Holt} \&
  E.~{Smith}, 163--175

\bibitem[{{Li} {et~al.}(2020){Li}, {Wang}, {Riechers}, {Walter}, {Decarli},
  {Venamans}, {Neri}, {Shao}, {Fan}, {Gao}, {Carilli}, {Omont}, {Cox},
  {Menten}, {Wagg}, {Bertoldi}, \& {Narayanan}}]{Li2020}
{Li}, J., {Wang}, R., {Riechers}, D., {et~al.} 2020, \apj, 889, 162

\bibitem[{{Liu} {et~al.}(2015){Liu}, {Gao}, {Isaak}, {Daddi}, {Yang}, {Lu}, \&
  {van der Werf}}]{Liu2015}
{Liu}, D., {Gao}, Y., {Isaak}, K., {et~al.} 2015, \apjl, 810, L14

\bibitem[{{Madau} \& {Dickinson}(2014)}]{Madau2014}
{Madau}, P. \& {Dickinson}, M. 2014, \araa, 52, 415

\bibitem[{{McMullin} {et~al.}(2007){McMullin}, {Waters}, {Schiebel}, {Young},
  \& {Golap}}]{McMullin2007}
{McMullin}, J.~P., {Waters}, B., {Schiebel}, D., {Young}, W., \& {Golap}, K.
  2007, in Astronomical Society of the Pacific Conference Series, Vol. 376,
  Astronomical Data Analysis Software and Systems XVI, ed. R.~A. {Shaw},
  F.~{Hill}, \& D.~J. {Bell}, 127

\bibitem[{{Meijerink} {et~al.}(2013){Meijerink}, {Kristensen}, {Wei{\ss}}, {van
  der Werf}, {Walter}, {Spaans}, {Loenen}, {Fischer}, {Israel}, {Isaak},
  {Papadopoulos}, {Aalto}, {Armus}, {Charmandaris}, {Dasyra}, {Diaz-Santos},
  {Evans}, {Gao}, {Gonz{\'a}lez-Alfonso}, {G{\"u}sten}, {Henkel}, {Kramer},
  {Lord}, {Mart{\'\i}n-Pintado}, {Naylor}, {Sanders}, {Smith}, {Spinoglio},
  {Stacey}, {Veilleux}, \& {Wiedner}}]{Meijerink2013}
{Meijerink}, R., {Kristensen}, L.~E., {Wei{\ss}}, A., {et~al.} 2013, \apjl,
  762, L16

\bibitem[{{Petter} {et~al.}(2023){Petter}, {Hickox}, {Alexander}, {Myers},
  {Geach}, {Whalen}, \& {Andonie}}]{Petter2023}
{Petter}, G.~C., {Hickox}, R.~C., {Alexander}, D.~M., {et~al.} 2023, arXiv
  e-prints, arXiv:2302.00690

\bibitem[{{Planck Collaboration} {et~al.}(2016){Planck Collaboration}, {Ade},
  {Aghanim}, {Arnaud}, {Ashdown}, {Aumont}, {Baccigalupi}, {Banday},
  {Barreiro}, {Bartlett}, {Bartolo}, {Battaner}, {Battye}, {Benabed},
  {Beno{\^\i}t}, {Benoit-L{\'e}vy}, {Bernard}, {Bersanelli}, {Bielewicz},
  {Bock}, {Bonaldi}, {Bonavera}, {Bond}, {Borrill}, {Bouchet}, {Boulanger},
  {Bucher}, {Burigana}, {Butler}, {Calabrese}, {Cardoso}, {Catalano},
  {Challinor}, {Chamballu}, {Chary}, {Chiang}, {Chluba}, {Christensen},
  {Church}, {Clements}, {Colombi}, {Colombo}, {Combet}, {Coulais}, {Crill},
  {Curto}, {Cuttaia}, {Danese}, {Davies}, {Davis}, {de Bernardis}, {de Rosa},
  {de Zotti}, {Delabrouille}, {D{\'e}sert}, {Di Valentino}, {Dickinson},
  {Diego}, {Dolag}, {Dole}, {Donzelli}, {Dor{\'e}}, {Douspis}, {Ducout},
  {Dunkley}, {Dupac}, {Efstathiou}, {Elsner}, {En{\ss}lin}, {Eriksen},
  {Farhang}, {Fergusson}, {Finelli}, {Forni}, {Frailis}, {Fraisse},
  {Franceschi}, {Frejsel}, {Galeotta}, {Galli}, {Ganga}, {Gauthier}, {Gerbino},
  {Ghosh}, {Giard}, {Giraud-H{\'e}raud}, {Giusarma}, {Gjerl{\o}w},
  {Gonz{\'a}lez-Nuevo}, {G{\'o}rski}, {Gratton}, {Gregorio}, {Gruppuso},
  {Gudmundsson}, {Hamann}, {Hansen}, {Hanson}, {Harrison}, {Helou},
  {Henrot-Versill{\'e}}, {Hern{\'a}ndez-Monteagudo}, {Herranz}, {Hildebrandt},
  {Hivon}, {Hobson}, {Holmes}, {Hornstrup}, {Hovest}, {Huang}, {Huffenberger},
  {Hurier}, {Jaffe}, {Jaffe}, {Jones}, {Juvela}, {Keih{\"a}nen}, {Keskitalo},
  {Kisner}, {Kneissl}, {Knoche}, {Knox}, {Kunz}, {Kurki-Suonio}, {Lagache},
  {L{\"a}hteenm{\"a}ki}, {Lamarre}, {Lasenby}, {Lattanzi}, {Lawrence}, {Leahy},
  {Leonardi}, {Lesgourgues}, {Levrier}, {Lewis}, {Liguori}, {Lilje},
  {Linden-V{\o}rnle}, {L{\'o}pez-Caniego}, {Lubin}, {Mac{\'\i}as-P{\'e}rez},
  {Maggio}, {Maino}, {Mandolesi}, {Mangilli}, {Marchini}, {Maris}, {Martin},
  {Martinelli}, {Mart{\'\i}nez-Gonz{\'a}lez}, {Masi}, {Matarrese}, {McGehee},
  {Meinhold}, {Melchiorri}, {Melin}, {Mendes}, {Mennella}, {Migliaccio},
  {Millea}, {Mitra}, {Miville-Desch{\^e}nes}, {Moneti}, {Montier}, {Morgante},
  {Mortlock}, {Moss}, {Munshi}, {Murphy}, {Naselsky}, {Nati}, {Natoli},
  {Netterfield}, {N{\o}rgaard-Nielsen}, {Noviello}, {Novikov}, {Novikov},
  {Oxborrow}, {Paci}, {Pagano}, {Pajot}, {Paladini}, {Paoletti}, {Partridge},
  {Pasian}, {Patanchon}, {Pearson}, {Perdereau}, {Perotto}, {Perrotta},
  {Pettorino}, {Piacentini}, {Piat}, {Pierpaoli}, {Pietrobon}, {Plaszczynski},
  {Pointecouteau}, {Polenta}, {Popa}, {Pratt}, {Pr{\'e}zeau}, {Prunet},
  {Puget}, {Rachen}, {Reach}, {Rebolo}, {Reinecke}, {Remazeilles}, {Renault},
  {Renzi}, {Ristorcelli}, {Rocha}, {Rosset}, {Rossetti}, {Roudier},
  {Rouill{\'e} d'Orfeuil}, {Rowan-Robinson}, {Rubi{\~n}o-Mart{\'\i}n},
  {Rusholme}, {Said}, {Salvatelli}, {Salvati}, {Sandri}, {Santos},
  {Savelainen}, {Savini}, {Scott}, {Seiffert}, {Serra}, {Shellard}, {Spencer},
  {Spinelli}, {Stolyarov}, {Stompor}, {Sudiwala}, {Sunyaev}, {Sutton},
  {Suur-Uski}, {Sygnet}, {Tauber}, {Terenzi}, {Toffolatti}, {Tomasi},
  {Tristram}, {Trombetti}, {Tucci}, {Tuovinen}, {T{\"u}rler}, {Umana},
  {Valenziano}, {Valiviita}, {Van Tent}, {Vielva}, {Villa}, {Wade}, {Wandelt},
  {Wehus}, {White}, {White}, {Wilkinson}, {Yvon}, {Zacchei}, \&
  {Zonca}}]{Planck2016}
{Planck Collaboration}, {Ade}, P.~A.~R., {Aghanim}, N., {et~al.} 2016, \aap,
  594, A13

\bibitem[{{Riechers} {et~al.}(2021){Riechers}, {Cooray}, {P{\'e}rez-Fournon},
  \& {Neri}}]{Riechers2021b}
{Riechers}, D.~A., {Cooray}, A., {P{\'e}rez-Fournon}, I., \& {Neri}, R. 2021,
  \apj, 913, 141

\bibitem[{{Rosenberg} {et~al.}(2015){Rosenberg}, {van der Werf}, {Aalto},
  {Armus}, {Charmandaris}, {D{\'\i}az-Santos}, {Evans}, {Fischer}, {Gao},
  {Gonz{\'a}lez-Alfonso}, {Greve}, {Harris}, {Henkel}, {Israel}, {Isaak},
  {Kramer}, {Meijerink}, {Naylor}, {Sanders}, {Smith}, {Spaans}, {Spinoglio},
  {Stacey}, {Veenendaal}, {Veilleux}, {Walter}, {Wei{\ss}}, {Wiedner}, {van der
  Wiel}, \& {Xilouris}}]{Rosenberg2015}
{Rosenberg}, M.~J.~F., {van der Werf}, P.~P., {Aalto}, S., {et~al.} 2015, \apj,
  801, 72

\bibitem[{{Schneider} {et~al.}(2015){Schneider}, {Bianchi}, {Valiante},
  {Risaliti}, \& {Salvadori}}]{Schneider2015}
{Schneider}, R., {Bianchi}, S., {Valiante}, R., {Risaliti}, G., \& {Salvadori},
  S. 2015, \aap, 579, A60

\bibitem[{{Scott} {et~al.}(2011){Scott}, {Lupu}, {Aguirre}, {Auld}, {Aussel},
  {Baker}, {Beelen}, {Bock}, {Bradford}, {Brisbin}, {Burgarella}, {Carpenter},
  {Chanial}, {Chapman}, {Clements}, {Conley}, {Cooray}, {Cox}, {Dowell},
  {Eales}, {Farrah}, {Franceschini}, {Frayer}, {Gavazzi}, {Glenn}, {Griffin},
  {Harris}, {Ibar}, {Ivison}, {Kamenetzky}, {Kim}, {Krips}, {Maloney},
  {Matsuhara}, {Mortier}, {Murphy}, {Naylor}, {Neri}, {Nguyen}, {Oliver},
  {Omont}, {Page}, {Papageorgiou}, {Pearson}, {P{\'e}rez-Fournon}, {Pohlen},
  {Rawlings}, {Raymond}, {Riechers}, {Rodighiero}, {Roseboom},
  {Rowan-Robinson}, {Scott}, {Seymour}, {Smith}, {Symeonidis}, {Tugwell},
  {Vaccari}, {Vieira}, {Vigroux}, {Wang}, {Wright}, \&
  {Zmuidzinas}}]{Scott2011}
{Scott}, K.~S., {Lupu}, R.~E., {Aguirre}, J.~E., {et~al.} 2011, \apj, 733, 29

\bibitem[{{Shao} {et~al.}(2019){Shao}, {Wang}, {Carilli}, {Wagg}, {Walter},
  {Li}, {Fan}, {Jiang}, {Riechers}, {Bertoldi}, {Strauss}, {Cox}, {Omont}, \&
  {Menten}}]{Shao2019}
{Shao}, Y., {Wang}, R., {Carilli}, C.~L., {et~al.} 2019, \apj, 876, 99

\bibitem[{{Shao} {et~al.}(2022){Shao}, {Wang}, {Weiss}, {Wagg}, {Carilli},
  {Strauss}, {Walter}, {Cox}, {Fan}, {Menten}, {Narayanan}, {Riechers},
  {Bertoldi}, {Omont}, \& {Jiang}}]{Shao2022}
{Shao}, Y., {Wang}, R., {Weiss}, A., {et~al.} 2022, \aap, 668, A121

\bibitem[{{Simpson} {et~al.}(2012){Simpson}, {Smail}, {Swinbank}, {Alexander},
  {Auld}, {Baes}, {Bonfield}, {Clements}, {Cooray}, {Coppin}, {Danielson},
  {Dariush}, {Dunne}, {de Zotti}, {Harrison}, {Hopwood}, {Hoyos}, {Ibar},
  {Ivison}, {Jarvis}, {Lapi}, {Maddox}, {Page}, {Riechers}, {Valiante}, \& {van
  der Werf}}]{Simpson2012}
{Simpson}, J.~M., {Smail}, I., {Swinbank}, A.~M., {et~al.} 2012, \mnras, 426,
  3201

\bibitem[{{Solomon} {et~al.}(1992){Solomon}, {Downes}, \&
  {Radford}}]{Solomon1992}
{Solomon}, P.~M., {Downes}, D., \& {Radford}, S.~J.~E. 1992, \apjl, 398, L29

\bibitem[{{Spilker} {et~al.}(2018){Spilker}, {Aravena}, {B{\'e}thermin},
  {Chapman}, {Chen}, {Cunningham}, {De Breuck}, {Dong}, {Gonzalez}, {Hayward},
  {Hezaveh}, {Litke}, {Ma}, {Malkan}, {Marrone}, {Miller}, {Morningstar},
  {Narayanan}, {Phadke}, {Sreevani}, {Stark}, {Vieira}, \&
  {Wei{\ss}}}]{Spilker2018}
{Spilker}, J.~S., {Aravena}, M., {B{\'e}thermin}, M., {et~al.} 2018, Science,
  361, 1016

\bibitem[{{Spilker} {et~al.}(2020){Spilker}, {Aravena}, {Phadke},
  {B{\'e}thermin}, {Chapman}, {Dong}, {Gonzalez}, {Hayward}, {Hezaveh},
  {Litke}, {Malkan}, {Marrone}, {Narayanan}, {Reuter}, {Vieira}, \&
  {Wei{\ss}}}]{Spilker2020b}
{Spilker}, J.~S., {Aravena}, M., {Phadke}, K.~A., {et~al.} 2020, \apj, 905, 86

\bibitem[{{Stacey} {et~al.}(2022){Stacey}, {Costa}, {McKean}, {Sharon},
  {Calistro Rivera}, {Glikman}, \& {van der Werf}}]{Stacey2022}
{Stacey}, H.~R., {Costa}, T., {McKean}, J.~P., {et~al.} 2022, \mnras, 517, 3377

\bibitem[{{Stacey} {et~al.}(2021){Stacey}, {McKean}, {Powell}, {Vegetti},
  {Rizzo}, {Spingola}, {Auger}, {Ivison}, \& {van der Werf}}]{Stacey2021}
{Stacey}, H.~R., {McKean}, J.~P., {Powell}, D.~M., {et~al.} 2021, \mnras, 500,
  3667

\bibitem[{{Stacey} {et~al.}(2018){Stacey}, {McKean}, {Robertson}, {Ivison},
  {Isaak}, {Schleicher}, {van der Werf}, {Baan}, {Berciano Alba}, {Garrett}, \&
  {Loenen}}]{Stacey2018}
{Stacey}, H.~R., {McKean}, J.~P., {Robertson}, N.~C., {et~al.} 2018, \mnras,
  476, 5075

\bibitem[{{Stanley} {et~al.}(2021){Stanley}, {Knudsen}, {Aalto}, {Fan},
  {Falstad}, \& {Humphreys}}]{Stanley2021}
{Stanley}, F., {Knudsen}, K.~K., {Aalto}, S., {et~al.} 2021, \aap, 646, A178

\bibitem[{{Travascio} {et~al.}(2020){Travascio}, {Zappacosta}, {Cantalupo},
  {Piconcelli}, {Arrigoni Battaia}, {Ginolfi}, {Bischetti}, {Vietri},
  {Bongiorno}, {D'Odorico}, {Duras}, {Feruglio}, {Vignali}, \&
  {Fiore}}]{Travascio2020}
{Travascio}, A., {Zappacosta}, L., {Cantalupo}, S., {et~al.} 2020, \aap, 635,
  A157

\bibitem[{{van der Werf} {et~al.}(2010){van der Werf}, {Isaak}, {Meijerink},
  {Spaans}, {Rykala}, {Fulton}, {Loenen}, {Walter}, {Wei{\ss}}, {Armus},
  {Fischer}, {Israel}, {Harris}, {Veilleux}, {Henkel}, {Savini}, {Lord},
  {Smith}, {Gonz{\'a}lez-Alfonso}, {Naylor}, {Aalto}, {Charmandaris}, {Dasyra},
  {Evans}, {Gao}, {Greve}, {G{\"u}sten}, {Kramer}, {Mart{\'\i}n-Pintado},
  {Mazzarella}, {Papadopoulos}, {Sanders}, {Spinoglio}, {Stacey}, {Vlahakis},
  {Wiedner}, \& {Xilouris}}]{vanderWerf2010}
{van der Werf}, P.~P., {Isaak}, K.~G., {Meijerink}, R., {et~al.} 2010, \aap,
  518, L42

\bibitem[{{Venemans} {et~al.}(2018){Venemans}, {Decarli}, {Walter},
  {Ba{\~n}ados}, {Bertoldi}, {Fan}, {Farina}, {Mazzucchelli}, {Riechers},
  {Rix}, {Wang}, \& {Yang}}]{Venemans2018}
{Venemans}, B.~P., {Decarli}, R., {Walter}, F., {et~al.} 2018, \apj, 866, 159

\bibitem[{{Venemans} {et~al.}(2020){Venemans}, {Walter}, {Neeleman}, {Novak},
  {Otter}, {Decarli}, {Ba{\~n}ados}, {Drake}, {Farina}, {Kaasinen},
  {Mazzucchelli}, {Carilli}, {Fan}, {Rix}, \& {Wang}}]{Venemans2020}
{Venemans}, B.~P., {Walter}, F., {Neeleman}, M., {et~al.} 2020, \apj, 904, 130

\end{thebibliography}

\begin{appendix} %First appendix
\section{ALMA Observation Details}\label{A1}
\begin{figure*}
       \centering
       \includegraphics[width=0.94\textwidth]{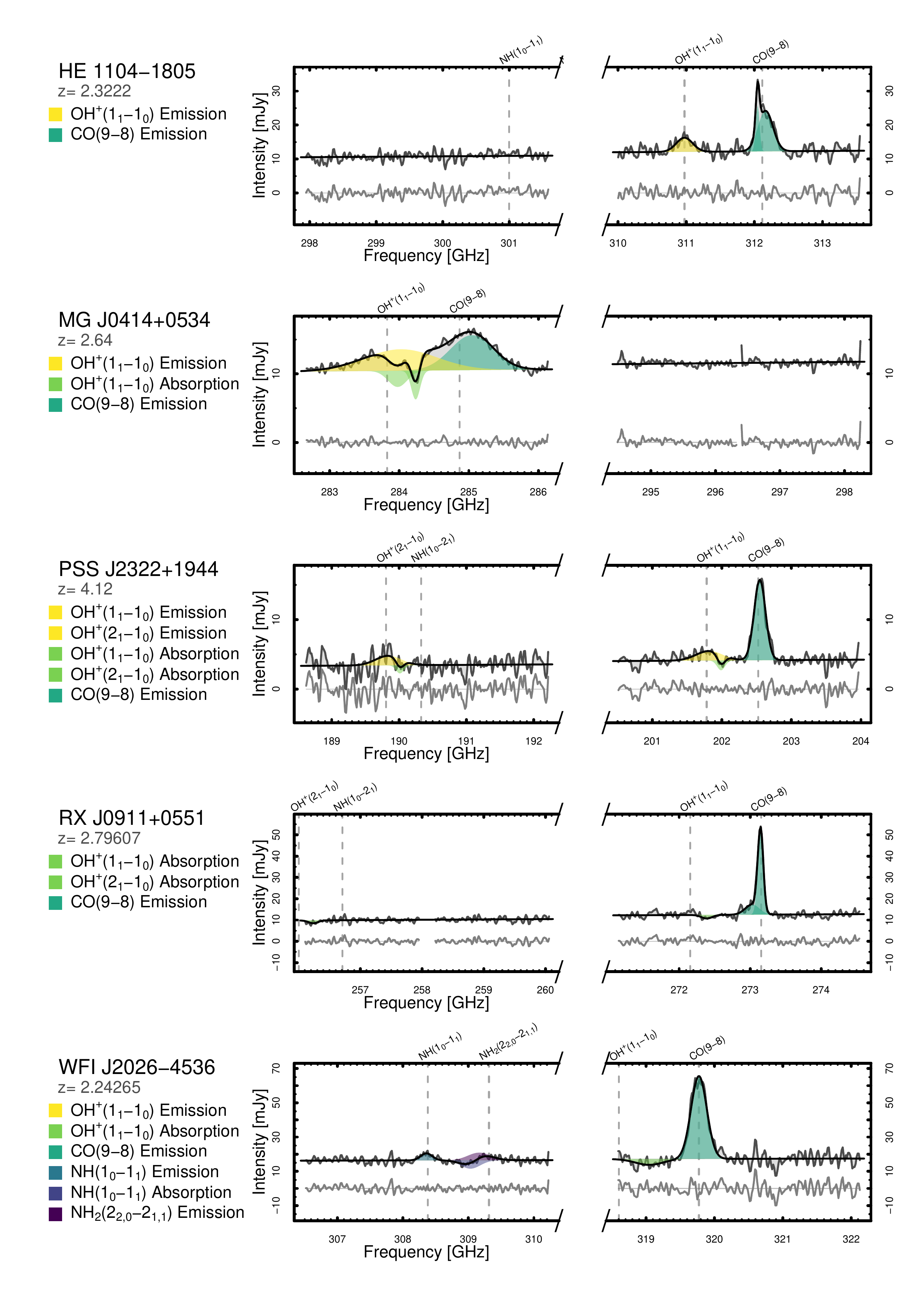}
       \caption{Spatially integrated spectra, fits and residuals of the 5 QSOs. Observed frequencies are displayed on the horizontal axis and the expected redshifted rest frequencies of the  spectral lines, using the redshifts in Table \ref{tab:derived}, are indicated by vertical dashed lines.\label{fig:fits}}
    \end{figure*}

    Here, we provide details of the ALMA observations in \ref{tab:obs} and data products in Fig. \ref{fig:fits}.
    
    \begin{table}[h!]
        \caption{ALMA observations}\label{tab:obs}
        \centering
        \begin{tabular}{lcccc} 
        \hline\hline             
        Name&$\rm Ant.s$ & TOS & Beam & $\sigma_{\rm 100kms^{-1}}$\\
         & & $\rm [s]$ & [$^{\prime\prime}\times^{\prime\prime}$] & $[\rm mJy\ beam^{-1}]$\\
        \hline
             MG J0414+0534 & $46$ & 3525 & $0.85\times0.96$&0.40\\
             RX J0911+0551 & $46$ & 1131 & $0.98\times1.2$&0.29\\
             HE 1104-1805  & $44$ & 2780 & $0.50\times0.65$&0.25\\
             WFI J2026-4536& $44$ & 1007 & $0.88\times1.0$&0.28\\
             PSS J2322+1944& $48$ & 3007 & $0.65\times0.90$&0.071\\
    \hline
    \end{tabular}
    \end{table}
    
\section{Additional lines in WFI J2026-4536}\label{A2}
We present the fitted and derived values of the spectral fits to the tentative detections of NH and $\rm NH_2$ in WFI~J2026-4536 in Table \ref{tab:obsWFI}. The tentaive $\rm NH(1_0-1_1)$ absorption and $\rm NH_2(2_{2,0}-2_{1,1})$ emission lines are blended, resulting in poorly constrained best-fit parameters for these lines. Flux, velocity, and velocity dispersion are all left as free parameters in the fitting of the three lines.

\begin{table}[h!]
    \caption{Observed and derived line properties of NH and $\rm NH_2$}\label{tab:obsWFI}
    \centering
    \begin{tabular}{llcc} 
    \hline\hline   
    \vspace{-3.5mm}\\
    \multicolumn{2}{l}{$\rm NH(1_0-1_1)$ Absorption} \\
    \hline
    $\rm S$&$\rm [mJy\ \kms]$& $-1.78\pm86$\\
    $\rm v$&[\kms]           & $-572\pm1100$\\
    $\rm \sigma$             &[\kms]& $145\pm560$\\
    $\rm \int\tau dv$ & [\kms]  & $94.8\pm4590$\\
    \hline
    \vspace{-3.5mm}\\
    \multicolumn{2}{l}{$\rm NH(1_0-1_1)$ Emission} \\
    \hline
    $\rm S$&$\rm [mJy\ \kms]$        & $0.890\pm0.15$\\
    $\rm v$&[\kms]                   & $81.3\pm15$\\
    $\rm \sigma$&[\kms]              & $86.5\pm15$\\
    $\rm\mu L$ &$\rm 10^{7}[L_{\odot}]$& $9.58\pm1.6$\\
    $\rm\mu L'$ &$\rm 10^{9}[L_{\odot}]$ & $2.97\pm0.49$\\
    \hline
    \vspace{-3.5mm}\\
    \multicolumn{2}{l}{$\rm NH_2(2_{2,0}-2_{1,1})$ Emission} \\
    \hline
    $\rm S$&$\rm [mJy\ \kms]$        & $1.90\pm86$\\
    $\rm v$&[\kms]                   & $204\pm4900$\\
    $\rm \sigma$&[\kms]              & $169\pm1000$\\
    $\rm\mu L$ &$\rm 10^{8}[L_{\odot}]$& 2.05$\pm93$\\
    $\rm\mu L'$ &$\rm 10^{9}[L_{\odot}]$ & 6.35$\pm288$\\
    \hline
\end{tabular}
\end{table}

\end{appendix}

\end{document}